\documentclass[]{article}

\usepackage{fullpage}

\usepackage{graphicx}        
\usepackage{multicol}        
\usepackage{multirow}

\begin{document}

\title{A Design Blueprint for Virtual Organizations in a Service Oriented Landscape}
\author{Wajeeha Khalil and Erich Schikuta\\
  University of Vienna, Faculty of Computer Science \\
  Research Group Workflow Systems and Technology \\
  W\"ahringerstr. 29, A-1090 Vienna, Austria\\
  email: erich.schikuta@univie.ac.at}


\date{}

\maketitle

\section*{Abstract}

``United we stand, divided we fall'' is a well known saying. We are living in the era of virtual collaborations. Advancement on conceptual and technological level has enhanced the way people communicate. Everything-as-a-Service once a dream, now becoming a reality.

Problem nature has also been changed over the time. Today, e-Collaborations are applied to all the domains possible. Extensive data and computing resources are in need and assistance  from human experts is also becoming essential. This puts a great responsibility on Information Technology (IT) researchers and developers to provide generic platforms where user can easily communicate and solve their problems. To realize this concept, distributed computing has offered many paradigms, e.g. cluster, grid, cloud computing. Virtual Organization (VO) is a logical orchestration of globally dispersed resources to achieve common goals.

Existing paradigms and technology are used to form Virtual Organization, but lack of standards remained a critical issue for last two decades. Our research endeavor focuses on developing a design blueprint for Virtual Organization building process. The proposed standardization process is a two phase activity. First phase provides requirement analysis and the second phase presents a Reference Architecture for Virtual Organization (RAVO). This form of standardization is chosen to accommodate both technological and paradigm shift. We categorize our efforts in two parts. First part consists of a pattern to identify the requirements and components of a Virtual Organization. Second part details a generic framework based on the concept of Everything-as-a-Service.

%

\section{Introduction}

``Resource/Service as a utility'' once a dream, is now a reality we are living with. Utility computing is not a new concept, but rather it has quite a long history. Among the earliest references is by John McCarthy\footnote{John McCarthy, speaking at the MIT Centennial in 1961, ``Architects of
the Information Society, Thirty-Five Years of the Laboratory for Computer
Science at MIT'', edited by Hal Abelson}.

Last two decades of Information Technology (IT) development has witnessed the  specific efforts done to make this statement of John McCarthy a reality. Utility computing is providing basics for the current day resource utilization. Cluster, grids and now cloud computing have made this vision a reality. E-Collaborations also called \emph{virtual organizations} have been evolved with the technological and paradigm shift. Cluster computing offered more centralized resource pool, while grid computing remaind in need via hardware and computation cycles offerings to the scientific community. Grid computing models observed a deadlock after the introduction of cloud computing concepts. This situation was a result of missing economic business models, although some work was done on this issue even by the authors~ \cite{weishaeupl_business_2005,schikuta_business_2005,weishaeupl_towards_2004,weishaeupl_gset:_2006}. Based on Pay-as-you-use criteria, cloud computing is still in early stage. However, the economic component of cloud computing is a central focus of actual research~\cite{cs3939,cs3713,machschikutaIIWAS12}. Research efforts are going on to establish the basis of cloud computing as Every-thing-as-a-Service paradigm.

Infrastructure providing resources as a utility must be dynamic, scalable and reliable. Orchestration of resources across the globe, named as Virtual Organization (VO)/Virtual Enterprize (VE) has been extensively deployed to achieve this target. Change in the hardware and software technology, computing paradigms algorithm and procedures, incorporation of knowledge rather information and data, made the concepts of VO vague. Though VO had been created utilizing the best technology known to that time, but the success was short lived. There are three main issues, which has to be considered in order to understand:
\begin{itemize}
\item{ Advancement in hardware/software technology.}
\item{ Birth of new computing paradigms.}
\item{ Changed nature of resources and requirements from end user.}
\end{itemize}
 We are living in the age of transformation. A paradigm shift is one that effects the society as a whole. According to Peter Drucker such transformation place over fifty to sixty-year periods \cite{Drucker1994}. In his book ``Post-Capitalist Society'', he outlines three earlier periods of dramatic changes in the Western World.
\begin{itemize}
\item The rise of medieval craft guilds and urban centuries. Long distance trade (thirteenth-Century Europe) \cite{Drucker1994}.
\item The renaissance period of Gutenburg's printing press and Lutheran Reformation (1455-1519) \cite{Drucker1994}.
\item The industrial revolution, starting with Watt's steam engine (1776-1815) \cite{Drucker1994}.
\end{itemize}

Existing technologies and paradigms do not vanish with the birth of new concepts rather they adopt what is positive and remove what is not required. Technology and paradigm used to form VO have also faced this transformation. For example networking, distributed computing, cluster computing, grid computing, utility computing and now cloud computing, all are related and are improvements of the existing concepts. When technology changes or improves, paradigm needs an upgrade too. New methods and algorithms are created to support the hardware. Another main factor is the requirements from the user community. The user community puts a demand on the technology and computing paradigm and they evolve accordingly.

``Resources/Services as a utility'' is main theme of collaboration. To achieve the goal(s), organizations and individuals gather all the resources available. The spectrum of availability has covered the whole globe. Today, time and space are not a limit due to Information and Communication Technology (ICT) advancements. This revolution has an impact on the resources types. Initial collaborations offered only storage and downloading (P2P networks), computing cycles and storage space (grid computing and cluster computing). Main focus remained at hardware and software sharing, but VOs for scientific research initiated another requirement, i.e. need of a human expert to guide the beginners in the said domain. Expert becomes an integral part of collaborations. Also, the two way contribution (duplex) motivated us to review and categorize the resource in the vicinity of VO. The categorization we presented is also vigilant to depict the general pattern of resources in any domain.

VO is the right place for both technology and computing paradigm to merge and achieve the objectives. In the past two decades collaborative computing has remained main concern of technology produced. Optimization of time and heterogeneous resources by building VO is the key point of today's research directions. Vision of a VO has evolved with the networking and distributed computing concepts.

Research community recognizes VO with different names, e.g. collaboratories \cite{Kesselman2008} \cite{WAWulf1993}, E-Science or E-Research \cite{Kesselman2008} \cite{Hey2005}, distributed work groups or virtual teams \cite{Kesselman2008} \cite{OLeary2007}, virtual environments, virtual enterprize \cite{Katzy2005} and online communities \cite{Kesselman2008} \cite{Preece2000}. Initially, focus was to improve business by utilizing the ability to gather resources which are scattered across the dimensions of time, space and structure. With the advent of modern technology, VO has encompassed almost all fields of life. We can say that every human will be soon part of a VO. VO's concepts need to be revisited with this evolution in general. VOs have been visioned from the business perspective in early 1990s. Pervasiveness of technology and improvements in computing paradigms has extended the domain of VO to cover all the areas where individuals and organizations meet to achieve some goal (formal Virtual organization) or without any specific common objective, e.g. social networks (informal VO). To the best of our knowledge, till now there are no standard procedures or patterns for how VO should be created and evolve to accommodate the changes in its integral parts or entities. Lack of standards for VO motivated us to provide a standard vision of E-collaboration incorporating both paradigm and technology shift as a Reference Architecture (RA) to achieve common objective(s) in any domain. Our research efforts also introduced new concepts regarding resources and stakeholder of a VO.

To provide a standard for VO, we consider the existing technologies and paradigms. Service Oriented Architecture (SOA), Web 2.0 and Web 3.0 are the underlying technological platform, and computing paradigms include utility computing and cloud computing.

During our research process we studied the existing infrastructures available for VO.
Utilizing electronic collaborations for achieving common goals is a tradition rather a requirement. Distributed resources are gathered using an infrastructure and are exploited to obtain the said results. In IT world such collaboration is known as VO. Idea is to provide resources as a utility to the end user. Service-Oriented infrastructures need to act dynamically to fulfil the demands from organizations and businesses. We encountered the following addressable issues:

\begin{itemize}
\item Does existing electronic collaboration approaches follow a standard?
\item Can we define patterns without predefined standards?
\item Does existing infrastructures fulfil the requirements of participating entities?
\item Are the existing infrastructures dynamic and adaptable to the rapidly updating IT and business world?
\item Can we design a generic platform to integrate resources from multiple domains using essential and optional parts?
\end{itemize}
	
Our research aims to answer these questions. VO's creation process lacks standards/patterns/methods \cite{Kesselman2008}. We analyzed existing VOs and the process of their creation through available documentation. We found the following answers to the above questions:

\begin{itemize}
\item Currently, there exist no specific standards for building VO or E-collaboration.
\item Existing infrastructures are modified for specific domain needs and cannot be generalized to all the domains. Since, existing technology is used without following any standard for creating a VO, it is hard to foresee incoming demands from the participants.
\item We require a generic platform to integrate resources from a single domain or multiple domains.
\item Defining a generic platform on the basis of Everything-as-a-Service (XaaS)\cite{Plummer2008cloud} concept is a solution.
Definition of participating components as:
	\begin{itemize}
		\item	Essential parts
		\item	Optional parts
	\end{itemize}
\end{itemize}

We present the following solutions for these obstacles:
\begin{itemize}
\item	Generalized patterns for building a VO.
\item	Defining components of a VO.
\item	Providing new definitions and examples of \emph{Resources} and \emph{Stakeholder} in different domains and justifying them in real world.
\item	Presenting a Reference Architecture for Virtual Organization (RAVO) which can be applied as a starting point for any community (belonging to a single or multiple domains) to collaborate.
\end{itemize}

The layout of the paper is as follows: In the next section \ref{sec:VO} we introduce the notion of Virtual Organization. In section \ref{sec:creatingVO} we present the principal steps for building a virtual organization, which we map to the process of our blueprint design. The generic reference architecture for virtual organizations, which we call RAVO, is laid out in section \ref{sec:ravo}.
For justification of our approach a proof-of-concept implementation based on RAVO is presented in section \ref{sec:n2sky}. Finally the paper is closed with a conclusion of the findings.

\section{Virtual Organization}
\label{sec:VO}
A Virtual Organization is a non-physical communication model with the purpose is of a common goal. It is built up from people and organizations with respect to geographical limits and nature.
Additionally, a Virtual Organization provides typically a collection of logical and physical resources distributed across the globe. From a conceptual point of view a \emph{Virtual Organization} resembles a detailed non-physical problems solving environment. Many definitions have been presented and many different terms arose, e.g. collaboratories \cite{Kesselman2008,WAWulf1993}, E-Science or E-Research \cite{Kesselman2008,Hey2005}, distributed work groups or virtual teams \cite{Kesselman2008,OLeary2007}, virtual environments, and online communities \cite{Kesselman2008,Preece2000}.
A Virtual Organization can comprise a group of individuals whose members and resources may be dispersed geographically and institutionally, yet who function as a coherent unit through the use of a Cyber Infrastructure \cite{Kesselman2008}. Virtual Organizations are typically enabled by and provide shared and, in most cases, real-time access to centralized or distributed resources, such as community specific tools, applications, data, instruments, and experimental operations, which was a key research areas of the authors~\cite{schikuta_vipios_2002,schikuta_vipios:_1998,brezany_software_1996} in the past. The different types of Virtual Organizations depend upon mode of operation, goals they achieve and life span for which they exists.

Regardless of the objectives, Virtual Organizations possess some common characteristics. Virtual Organizations provide distributed access across space and time. Structures and processes running a Virtual Organization are dynamic. Email, video conferencing, telepresence, awareness, social computing and group management tools are used to enable collaboration among the participants \cite{Kesselman2008}. Operational organizations are supported by simulation, database and analytical services. In daily life we come across many Virtual Organizations in terms of social networks as well (e.g. Facebook, MySpace). It can be phrased that soon every human on this earth will considered to be a part of some Virtual Organization serving its purpose in the said organization.

Generally, a Virtual Organization provides a global problem solving platform. It is difficult to specify or restrict the domain for which they are serving. Some advantageous roles played by Virtual Organizations are facilitator of access (BIRN \cite{birn}, LEAD \cite{lead}, nanoHUB \cite{nanohub}), caBig\cite{cabig}, LHC Computing Grid, enhancer of Problem Solving Processes (TeraGrid \cite{teragrid}) and  key to Competitiveness (GEON \cite{geon}). Virtual Organizations have served in the field of earthquake engineering (Southern California Earthquake Center (SCEC)), cancer research (Cancer Biomedical Informatics Grid (caBIG) \cite{cabig}), climate research (Earth System Grid \cite{earthsystemgrid}), high-energy physics (Large Hadron Collider), and computer science. Other communities are now forming Virtual Organizations to study system-level science. These Virtual Organizations are addressing problems that are too large and complex for any individual or institution to tackle alone. It simply is not possible to assemble at a single location all of the expertise required to design a modern accelerator, to understand cancer, or to predict the likelihood of future earthquakes. Virtual Organizations allow humanity to tackle previously intractable problems \cite{Kesselman2008}.


\section{Building a Virtual Organization}\label{sec:creatingVO}
The creation of a Virtual Organization is time consuming and should be a well planned activity. In this section we will discuss Virtual Organization and technology from different perspectives. Both aspects are required to support each other. Technology provides the basic infrastructure for a Virtual Organization to exist. A Virtual Organization in turn places demands on Information Technology and shapes the evolution of technology. For the last decade the Virtual Organization is one of the most discussed collaboration environment; but still no standards exist.
From this discussions we assume a step wise approach which is helpful in the creation of Virtual Organization. It can be separated in two phases which are detailed below.

\subsection{Phase-1: Questions}
The definition of a Virtual Organization starts with a series of questions, which are very critical in order to proceed. These questions (Qx) are listed in the following:

\begin{itemize}
\item Q1: Why to form a Virtual Organization? What are the reasons of an organization to create a Virtual Organization?
\item Q2: What is the motivation behind participation? Why should other persons, institutes, service providers, etc. want to participate in a Virtual Organization?
\item Q3: What services are offered by a Virtual Organization?
\item Q4: How are these services fared? What is the type of the resources/business model?
\item Q5: Who are the intended users? Who will eventually use and get benefited from this Virtual Organization?
\item Q6: What is the life of (membership of) a Virtual Organization? Are temporal alliance or permanent participation expected?
\end{itemize}

\subsection{Phase-2: Identification of Components}
Based on these Q\&A activity it is necessary to identify the building blocks of a Virtual Organization. Gannon \cite{Gannon2008} has identified main components of a Virtual Organization. These components are

\begin{itemize}
\item \emph{Common interest.} The reason to form a Virtual Organization,
\item \emph{Users.} the participants of a Virtual Organization,
\item \emph{Tools and services.} This is a crucial part of a Virtual Organization, which maintains the overall working environment and saves the existing patterns to be reused in order to reduce time to solve similar problems. A Virtual Organization requires a collection of shared analysis tools (e.g. visualization tools and provenance tools). Tools can be integrated into specific Virtual Organization work flows and can be shared and reused. They are used to collect data and publish results.
\item \emph{Data.} A Virtual Organization contains two types of data, generally categorized as meta data and operational data that is being operated by tools.
\end{itemize}

The component identification process provides the basic building blocks. The designer of a Virtual Organization can decide what to be improved and further included in the design process. Also, each component can be given a unique definition by the designer in context of a Virtual Organization being created.  Detailed information about creation, management and application area of Virtual Organizations is available in \cite{Kesselman2008}.

\section{RAVO: A Reference Architecture for Virtual Organizations}\label{sec:ravo}
According to Gerrit Muller \cite{Gerrit} there are two simultaneous trends,
\begin{itemize}
\item Increasing complexity, scope and size of the system of interest, its context and the organizations creating system \cite{Gerrit}.
\item Increasing dynamics and integration: shorter time market, more interoperability, rapid changes and adaption in the field, in a highly competitive market, for example cost and performance pressure  \cite{Gerrit}.
\end{itemize}
These trends form basis for our proposed RA as well. VOs are developed as distributed system at multiple locations, by multiple entities, consisting of multiple applications by multiple vendors, merging multiple domains for providing solutions to multiple problems. RA comes in scene where the multiplicity reaches a critical mass triggering a need to facilitate product creation and life cycle support in this distributed open world \cite{Gerrit}.
We detail the RAVO in the subsequent sections.
We define RAVO as ``an open source template that does not only depict the architectural patterns and terminology, but also defines the boundaries where heterogeneous resources from different domains merge collaboratively into a common framework''.
A RA has a life span and is dependent on the target architecture and possibly other RAs. As guideline for our effort we closely analyzed the RA presented by SHAMAN (European Commission, ICT-216736),  GERRIT MULLER \cite{Gerrit} and NEXOF\cite{nexof}. RAVO provides
\begin{itemize}
\item A common lexicon and taxonomy.
\item A common (Architectural) vision.
\item Modularization and complementary context.
\item A layered approach(bottom-up).
\end{itemize}

\subsection{Goal}
A common vision facilitates the participating entities to work as a team to achieve their decided goals. Modularization helps to integrate different domains thereby decreasing the efforts and context information make the dynamic nature of the architecture consistent.

We aim for developing a RA which allows for new forms of IT infrastructure coping with new collaborative processing paradigms, as grid computing and cloud computing. Thus we have to deliver an environment to allow for the new \emph{Internet of Services and Things} accommodating the novel service stack, as IaaS, PaaS and SaaS.
Architecture is classified into different layers according to the service each layer provides. Layered architecture is chosen because it helps to group different components (logical and physical) according to the degree of relatedness and required functionality.

\subsection{Components and SPI based Framework}
RAVO is based on SPI model. Layered approach is used to achieve the goal of providing all the resources as a service. Layers are distributed into 3 broad categories, IaaS, PaaS, SaaS

\begin{figure}
  \begin{center}
  \includegraphics[width=5in]{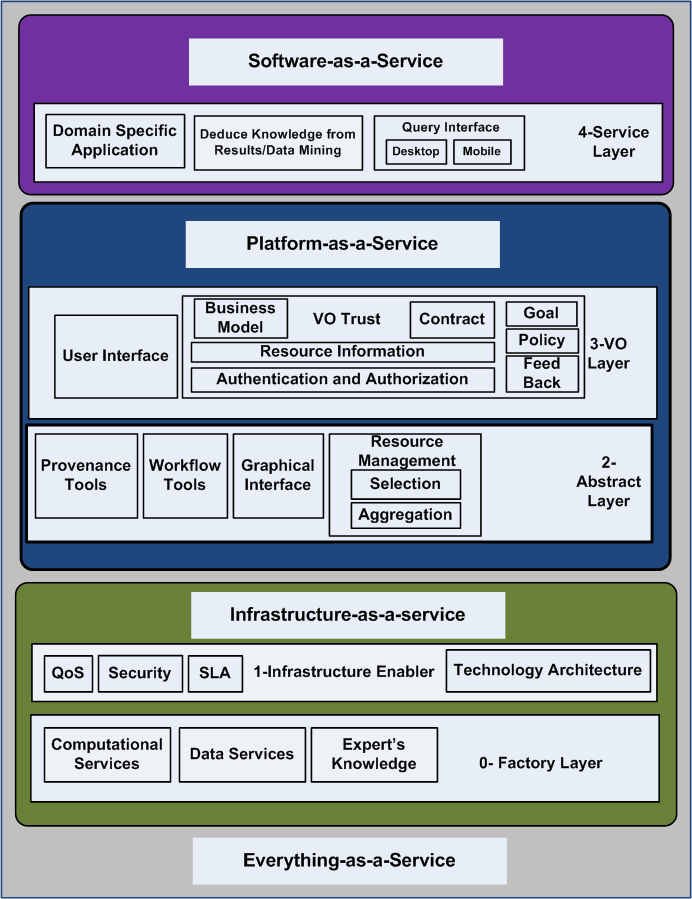}
  \caption{XaaS Skeleton of RAVO}
  \label{Fig:CSVO}
  \end{center}
   \end{figure}

Figure \ref{Fig:CSVO} presents the framework for VO using the SPI model. The layers are distributed into 3 broad categories, IaaS, PaaS and SaaS thus resulting in XaaS.

\subsubsection{\large{Software as a Service Layer}}

In context of RAVO, SaaS is composed of a Service layer. It contains Domain Specific Applications (DSA) accessible by all users. DSAs are the combination of several user interfaces and Business Models found in the VO layer. Users, who only use the platform to solve their domain specific problems and do not contribute to the VO, find an entry point at this layer.
\begin{itemize}
\item
\emph{Service Layer}: It has open source, downloadable software, categorized in domains. The Service layer packages several services provided by the VO layer to be subscribable entities. These entities include generic functionality to query information from the problem domain as well as the means to perform data mining on the compound data created or provided by the combination of the services.
\end{itemize}
\subsubsection{\large{Platform as a Service Layer}}

In RAVO two layers, namely VO layer and Abstract layer, cover PaaS.
\begin{itemize}
\item
\emph{Abstract Layer}: This layer is composed of essential tools which enable the whole framework to be exploited in a dynamic manner. The set of tools consist of provenance, workflow, graphical tools and any other domain specific tools which are used to enhance the reuse of the resources for a diverse set of problem solving activity. Each tool provides its own functionality, its own user interface description \cite{Altafwcci}, as well as an abstract API (identical for each tool) to access the resource in Factory layer.
\item \emph{VO Layer}: This layer is the entry point for user. It provides the realization of the user interface description and defines a business model on top of the Abstract layer to set usage cost according to usage statistics. Participating entities can agree on a usage model and build a cost trust for selling their resources. In context of VO, contributor/subject users (who not only use the resources offered by a VO but also contribute to the VO) are authorized to access the system on this layer. All have access to the system on PaaS layer.
\end{itemize}

\subsubsection{\large{Infrastructure as a Service Layer}}

 In RAVO logical and physical resources are considered to be the part of IaaS. This part consists of two sub-layers in RAVO: Factory layer and Infrastructure Enabler layer. Only users with administrative rights have access to this layer.
\begin{itemize}
\item \emph{Factory Layer}: Belongs to the IaaS category and contains resources for  RAVO. Resources are described as physical and logical resources. Physical resources comprise of hardware devices for storage and computation cycles in a distributed manner. Logical resources contain expert's knowledge that supports the problem solution activity thereby reducing time to reach the specified goal.
\item \emph{Infrastructure Enabler Layer}: Allows access to the resources provided by the Factory layer. It consists of protocols, procedures and methods to contact the desired resources for a problem solving activity. It acts as a glue or medium to reach the desired resources based on user request.
\end{itemize}

\subsubsection{\large{Everything as a Service Layer}}

All layers are providing their functionality in a pure Service-Oriented manner so we can say that RAVO is XaaS.

\subsection{Design Perspective of RAVO}
VOs is a broad category of distributed systems. It is envisioned as a combined effort of multiple entities (organizations, people, HW, SW) for achieving a goal. Building RA for VO is effective in many ways. RAVO forms basis for VOs belonging to any domain. It improves the effectiveness by managing synergy, providing guidance for collaboration, generic framework, managing and sharing the architectural patterns.
Interoperability is the most critical aspect of collaborative computing and VOs main feature. It determines the usability, performance and dependability of user level applications \cite{Gerrit}. Integration cost and time are also important factors in context of interoperability. RAVO supports interoperability by defining a negotiation model/trust for the participating entities thereby supporting the effective re-use of patterns.
Many RA focuses the technical architecture only. According to SAF meeting conclusion, A RA should address \cite{Gerrit},
\begin{itemize}
\item Technical architecture.
\item Business architecture.
\item Customer context.
\end{itemize}

RAVO well addresses these three aspects. It presents a technical architecture specifying the must participating modules, APIs, protocols and platform to support VO. RAVO offers a Business Model which is open according to the participating entities conditions for resource sharing. Business Model and customer context overlap. RAVO explicitly defines roles of participating entities as \emph{Subject}, \emph{Consumer}, \emph{Producer} and \emph{administrators}. Elaboration of roles makes it easy to dynamically update the Business Model as an entity changes the role. RAVO supports feedback from the participating entities which is helpful in improving and maintaining the existing RA. These concepts are already detailed in RAVO section.

RA is a perceived image of existing technologies. Designing RA is a challenging job because it needs sufficient proof to justify its need in the said context. RAVO focuses on VOs. To the best of our knowledge, there is no standard pattern or framework which can be used to create a VO from scratch. Our vision is to provide the VO community a complete framework for identifying main components and abstract a life cycle to create VO from scratch. It grasps knowledge from existing structures such as NEXOF \cite{nexof} and SHAMANs (European Commission,ICT-216736). Guidelines are used to modify the requirements into an RA which supports creation, dynamic evolution and maintenance of a VO.

\subsection{Viewpoint}
\begin{figure}
  \begin{center}
\includegraphics[width=4in]{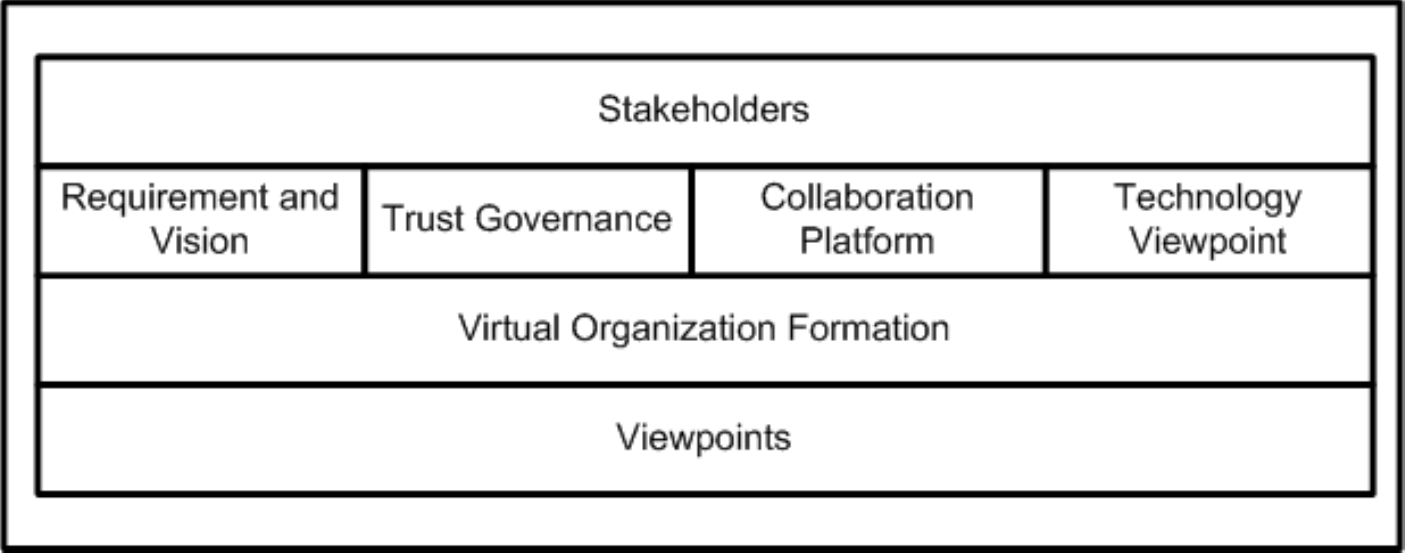}
  \caption{Viewpoints in Reference Architecture for Virtual Organization}
  \label{fig:VP}
  \end{center}
\end{figure}

Viewpoint is defined as a specification of the conventions for constructing and using a view. A pattern or template from which to develop individual views by establishing the purposes and audience for a view and the techniques for its creation and analysis \cite{Jen2000}. View is a representation or description of the entire system from a single perspective. Stakeholder is the viewer, who perceives the system according to her role. Viewpoint has a name, stakeholders addressed by it and concerns to be addressed by the viewpoint, and the language, modeling techniques or analytical methods to be used in constructing a view  based on the viewpoint \cite{Jen2000}.
According to these definitions, viewpoints extracted from the concerns of the stakeholders are shown in Figure \ref{fig:VP}. These viewpoints are detailed in the following sections.

\subsubsection{Forming a Virtual Organization} Stakeholders collaborate to form a VO. All participants of a VO have an objective (personal or organizational) to achieve via this collaboration. Sub-viewpoints are,
    \begin{itemize}
    \item Domain definition: Depends on the type of problem solution, target domain can be one or multiple. Thus, stakeholders can be from one or multiple domains.
    \item Participation Level: Participation can be individual or at organizational.
    \item Duration: Stakeholder remain part of the VO according to the membership duration agreed upon among the collaborative entities. It can vary depending on the type of VO, partial or permanent (either participation is required for a specific part or throughout) and a Business Model in case of profitable organizations.
    \item Types of Contribution: It is decided by the role assigned to a specific stakeholder in the context of a VO.
    \end{itemize}

\subsubsection{Requirements and Vision} This viewpoint formulates the boundaries of a VO and its participants. All the participants must clearly put their requirement and goals while building collaboration. These requirements should reflect any assumption made on the architecture and the respective requirements stemming from the assumptions. Once requirements are defined (in form of a list, catalogue), VO has a vision to achieve and targets are set accordingly.

\subsubsection{Trust Governance} Trust governance viewpoint is very important to any collaboration specially for VOs. Keeping stakeholders and resources glued together to achieve a target is only achievable via strong trust. Following sub-viewpoints are defined in this context:
\begin{itemize}
\item Trust/Policy formation:  Experts and planners from participating entities prepare an agreed upon policy/model/contract. This policy defines the rules for participating and leaving VO, contributing and  consuming resources, penalties for violation and measures to keep consistent and just to all the stakeholders.
\item Objective Catalogue: This viewpoint provides the list of all the contracts and agreements in a documented form necessary for authentication, authorization and stakeholder management.
\item Reviewing Policy: Due to dynamic nature of VO, the policies and contracts are reviewed to be in the accordance of change in requirements, technological updates, removal and entrance of participants.
\item Business Perspective: This viewpoint is optional depending upon the type of VO. Profitable VO have a Business Model for metering, billing in addition to authentication and  user management.
\end{itemize}

\subsubsection{Collaboration Platform } This view point provides details of participating components for realizing the VO on technological and system level. It is further divided into 4 sub-viewpoints which are briefed here as,
\begin{itemize}
\item Data: This viewpoint aims to depict the types of data utilized in collaboration. Two broad identifications are found as \emph{meta data} and \emph{operational data}. Problem nature, domain and participating entities decide on the data source and security in collaborative efforts. Data and relationship between different components can be represented using  Table, Mat, UML Class diagrams, Activity diagrams and Component diagrams.
\item Applications and Tools: This viewpoint describes the list of running applications and tools utilized in a problem solving activity. This view can be further divided according to requirement.  Roles of stakeholders also decide the access to different available tools and applications at multiple levels (Interface, infrastructure, platform and so on). Distribution and relationship among applications, tools and components can be shown using UML Component diagram.
\item Resources: This viewpoint explains the list of resources (Table, List), their owners, availability, usage cost (in case of profitable organization) and access rights. We have to sub-viewpoints:
    \begin{itemize}
    \item Subject: An important viewpoint which defines stakeholder which consume and contribute to the resources simultaneously.
    \item Enabler: This viewpoint details the stakholders which are related to deployment, configuration, monitoring and lifestyle management. Roles assigned in this viewpoint are developers, administrators, business providers, planners and experts.
\end{itemize}
\item Log catalogs: This viewpoint keep track of activities which are carried out during problem solving activities. Dynamic collaboration environments need to this record for the feedback and  improvement.
        \end{itemize}

\subsubsection{Technology Viewpoint}
This viewpoint lists the best available technology currently deployed. If new technology is employed which is not listed then it should be added to the list later. It is very helpful keeping VO consistent with the upcoming demands from business and user requirements and advancement in new computing paradigms and methods. Platforms used for collaboration have remained in a constant up-gradation. Choice must be made on technology by giving weighage to QoS, security, cost effective and timely solution to the end user.
An important sub-viewpoint of technological aspect is virtualization. It provides the way to reuse hardware cost, respond dynamically and maximize resource utilization and easy relocation. Virtualization viewpoint  deals with logical resources rather than physical resources.

All these viewpoints are shown in the diagram Figure \ref{fig:VP}
These viewpoints can be represented using Lists, Tables, UML tools, and other requirement specification tools available. They are also extendable and organizations can add any further categories according to their goals.

\subsection{Interface Description of Components}
RAVO is composed of multiple layers and each layer provides a set of components which are the building block of a VO. Selection of these building blocks is subjected to various aspects (i.e. life span, nature (dynamic or static), type, formal, informal and so on). We define interfaces for these components by specifying parameters (mandatory and optional), methods and necessary conditions for their executions.

\subsubsection{ VO Specifications }
VO needs to keep specific information in general, when created. It possesses some characteristics, (e.g. Unique ID, Date Created, Description about purpose domain etc). It also requires to maintain information about participating organizations and individuals.

\subsubsection{Resource Provider Information}
It is a must to maintain and update the information about the resource providers in a VO. Organization offering resources, time period for which resources are made available and access rights are potential characteristics.
\paragraph{Query Interface.}
RAVO proposes Query Interface as a mandatory component at Service Layer. User is facilitated with remote or desktop access. Query Interface enables user to search for their problem solution in Knowledge base. Knowledge base contains history of problems solved previously. On successful query user is provided with appropriate output. In case of no matching solution, query is processed and problem solutions is provided to user and Knowledge base is updated. Query Interface must provide login facility, identify the query type, check for existing solutions and must maintain a tolerable response time.


\paragraph{Domain Specific Application (DSA).}
DSA is a mandatory component DSA provide user with the ability to either download the applications and run on their own systems or use them at VO platform for problem solution. The range of applications depends on the domain and type of VO. Stakeholder can share their applications paid or non-paid basis. Sharing of application can be conditional (e.g. fully or partially paid in case of profitable organization). Information maintained about DSA must include how it is accessible (online or offline), access rights and cost (as defined in the Contract/Business Model).
\paragraph{Data Mining Tools.}
Data mining tools are an optional component of RAVO. They are a must for analytical and scientific research based VOs. Interface for data mining tools include tool description , access rights and  manul/help.

\subsubsection{PaaS Layer}
PaaS layer is composed of two layers, namely 3-VO Layer and 2-Abstract Layer. Component Specification is detailed below.
\paragraph{VO Trust.}
VO Layer consists of two main mandatory components. VO Trust is the most important of all components. It is formed by combining different modules and performs multiple tasks. It is responsible for \emph{Authentication} and \emph{Authorization} of VO members. Authorization is done on the basis of \emph{Roles} defined in the \emph{Contract/Business Model}. VO Trust have a mandatory emph{Contract} which consists of policies to achieve the goals of VO. In profitable or partially profitable VOs Business Model is also mandatory component of VO trust. In RAVO Business Model is optional and depends on the type of VO, however Contract is mandatory. Access rights are defined in contract or Business Model. Different methods are available to define the access rights. Organization models and access rights are comprehended in \cite{Rinder2008}. According to the authors access rights might be subjected to \emph{Organizational} and \emph{Direct} change \cite{Rinder2008}.  All components of VO Trust are synchronized to maintain the VO. Each component is assigned a specific task and output of one component provides input to the other component.  VO Trust has a \emph{Resource Information} component that acts like a \emph{Registry}. It keeps necessary details about all the resources available in VO.
\paragraph{User Interface.}
User Interface is a mandatory component of VO Layer. It provides access to the platform services offered by VO. User is authenticated and authorized using Login option. After authorization, user can formulate different queries and perform actions. These facilities are realized using a Web portal.
\paragraph{Workflow Tools.}
Abstract Layer is a sub layer of PaaS layer. It includes different components. Workflow tools is a mandatory component of this layer. Workflow management is a critical aspect of a VO in any domain. It supports Provenance management which plays vital role in monitoring and maintaining a VO. Workflow can be interpreted in different forms (e.g. graphical, textual, source code). Interpretation mode is chosen on the level of audience a VO possess. Workflow tools keep track of all the processes active in VO. Process management can be included as a sub component of a Workflow Tools. Dynamic adaption of in-process workflow is an essential part of any workflow management system. Classification of approaches along their strength and limitations used for dynamic adaption in workflow systems are detailed in \cite{Rinder2004}. Flexibility criteria in process management to handle the foreseen and unforeseen behaviors are categorized in \cite{FProcess2008}.

 Workflow tools allow user to define workflows for a problem solving activity. The participants responsible at each stage of this activity are notified and are responsible for delivering the promised results. Workflows are reusable and reduce redundancy and time for similar problems. Information maintained consist of workflow ID, type, status, access rights, how it interprets the results and process management. workflows are used by Provenance management to track the problem solving activity on user demand.

\paragraph{Provenance Tools.}
With the advent of financial computing
systems, as well as of data-intensive scientific collaborations, the
source of data items, and the computations performed during the
incident processing workflows have gained increasing importance \cite{Hasan2007}. Provenance of a resource is a record that
describes entities and processes involved in producing and delivering or otherwise influencing that resource\footnote. In a VO, provenance forms a critical foundation for enabling trust, reproduction and autentication.
Provenance assertions are a form of contextual metadata and can themselves
become important records with their own provenance. Provenance Tools are mandatory and included in Abstract Layer of RAVO. Provenance management is dependent on authorization, query management and  workflow management.

\paragraph{Graphical Interface.}
Graphical Interface is a mandatory component of Abstract Layer. It facilitates users to perform different task in VO Web portal. It provides an understandable interface to interact with the VO.

\paragraph{Resource Management.}
Resource Management is a mandatory component of Abstract Layer. It provides a mechanism to select and aggregate resources for a problem solving activity. Depending upon the underlying technology, VO developers can deploy different resource management tools. Necessary information maintained depends on the resource type and interest of participating entities. Basic information includes resource's unique identification, categorization as logical or physical, owner information, access rights and costs etc.

\subsubsection{IaaS Layer}
IaaS layer is composed of Infrastructure Enabler Layer and Factory Layer. This layer from the fabric of RAVO. All the resources are avaialble in Factory Layer and are exploited through Infrastructure Enabler Layer.
\paragraph{Infrastructure Enabler.}
This module is depending on the underlying technology . QoS, Service Level Agreement (SLA), Security, Fault tolerance and Disaster management are  most important issues specifically in clouds~\cite{haq_sla_2010,haq_aggregation_2010,haq_sla_2010-1,haq_conceptual_2009,haq_rule-based_2009,haq_aggregating_2009}. These aspects have  to be implemented on the bases of terms and conditions presented by participating entities. Financial aspect is another limitation for the implementation of these modules. Any other desired aspects can be added to extend the Infrastructure enabler layer. Components are dependent on the decision of the developers. RAVO identifies least basic and gives developers an open end to use them as mandatory or optional in their target VO.

\paragraph{Resource Catalogue.}
This module is part of Factory Layer but not explicitly shown in RAVO. It acts like a database for the resources. VO developers can include it at any layer according to their needs. RAVO keeps it at the Factory Layer as a  mandatory component. It contains information about resource management.

\paragraph{Expert.}
Expert represents the logical resource in RAVO Factory Layer. An Expert plays important role in problem solving activity. Expert can be contacted online during the problem solving process or she can be accessed offline. VO must maintain detailed information about Expert so that this feature can be fully exploited.

\paragraph{Data Service.}
Data Services is a mandatory component of Factory Layer. It represents the physical resource in RAVO. Data stores are important scientific and research based VOs.

\paragraph{Computational Services.}
Computational Services are mandatory component of Factory Layer. They also form the physical resources offered by a VO.

\section{Proof-of-Concept: The N2Sky System}\label{sec:n2sky}


As proof-of-concept of our approach we used RAVO as a design blueprint for implementing a cloud based VO for Neural Network Research, namely, N2Sky. We based the development of N2Sky on the blueprint provided by RAVO and  produced a concrete instance out of our proposed standard \cite{mann13}. This section compares N2Sky with RAVO to reveal the process of creation of N2Sky. The comparison justifies and proves how RAVO supported different development phases of N2Sky. We explain N2Sky as an instantiation of RAVO but with concrete components. We divide this comparison in 3 levels. First, \emph{Requirement Analysis Phase} that defined boundaries of N2Sky. Second, \emph{Component Identification Phase} which made it easy to identify the components of N2Sky and also choose between optional and mandatory components. Third, \emph{Implementation Phase} that reveals how technology independence, XaaS and layered distribution of components made it helpful to implement the system. The stakeholders envisioned in RAVO are also implemented as part of N2Sky.

\subsection{Requirement Analysis in terms of RAVO}
In the previous section we detailed a series of questions which must be answered by the responsible authorities for creating a VO. N2Sky utilizes this pattern for defining the requirements boundary of the system. These questions are answered in detail in an interview by engaged software engineering experts, whic are presented in section \ref{sec:evaluation}.
\subsection{ Component Identification in terms of RAVO}
\begin{figure}
  \begin{center}
  \includegraphics[width=5in]{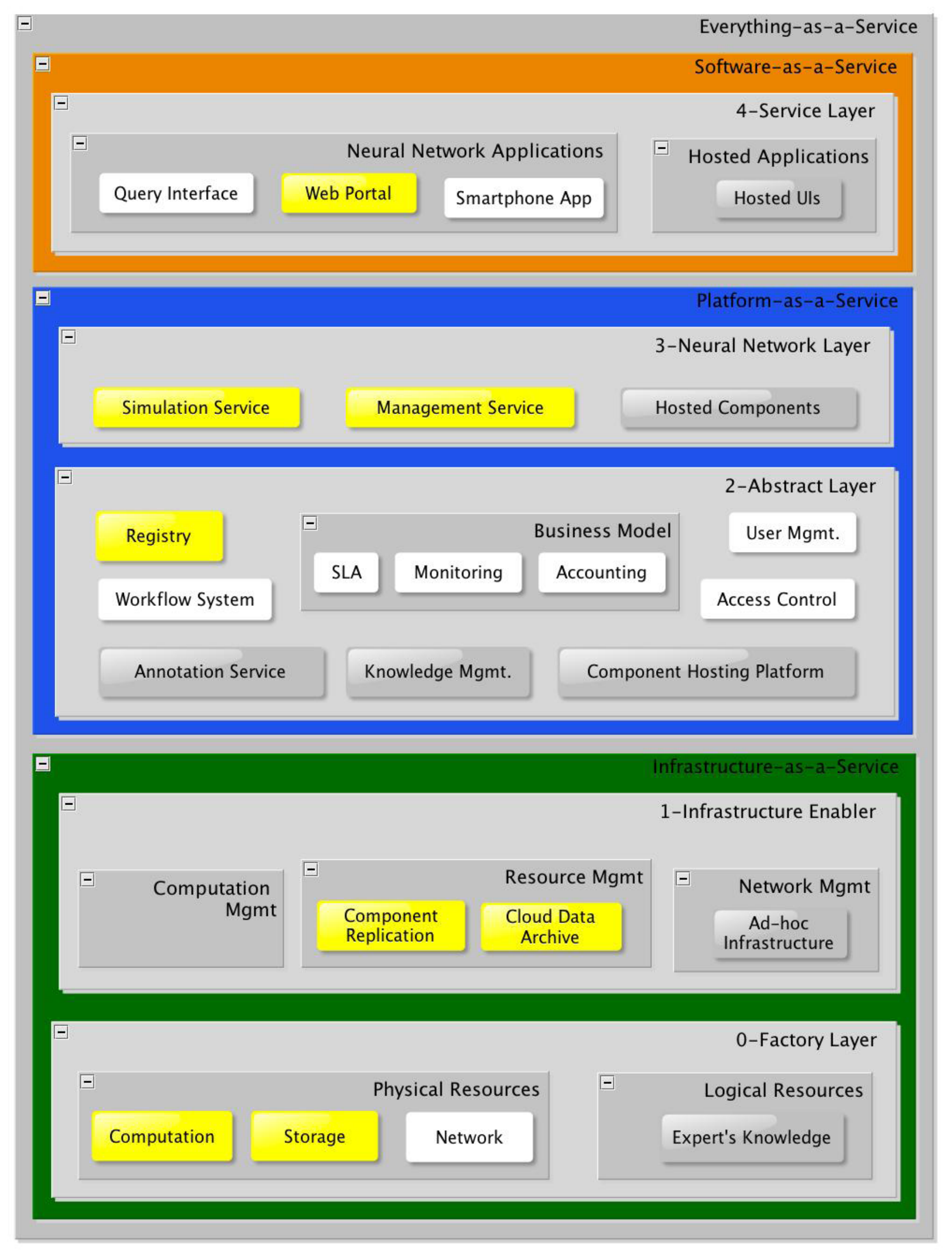}
  \caption{N2Sky}
  \label{Fig:N2S}
  \end{center}
   \end{figure}

\begin{figure}
  \begin{center}
  \includegraphics[width=0.8\textwidth]{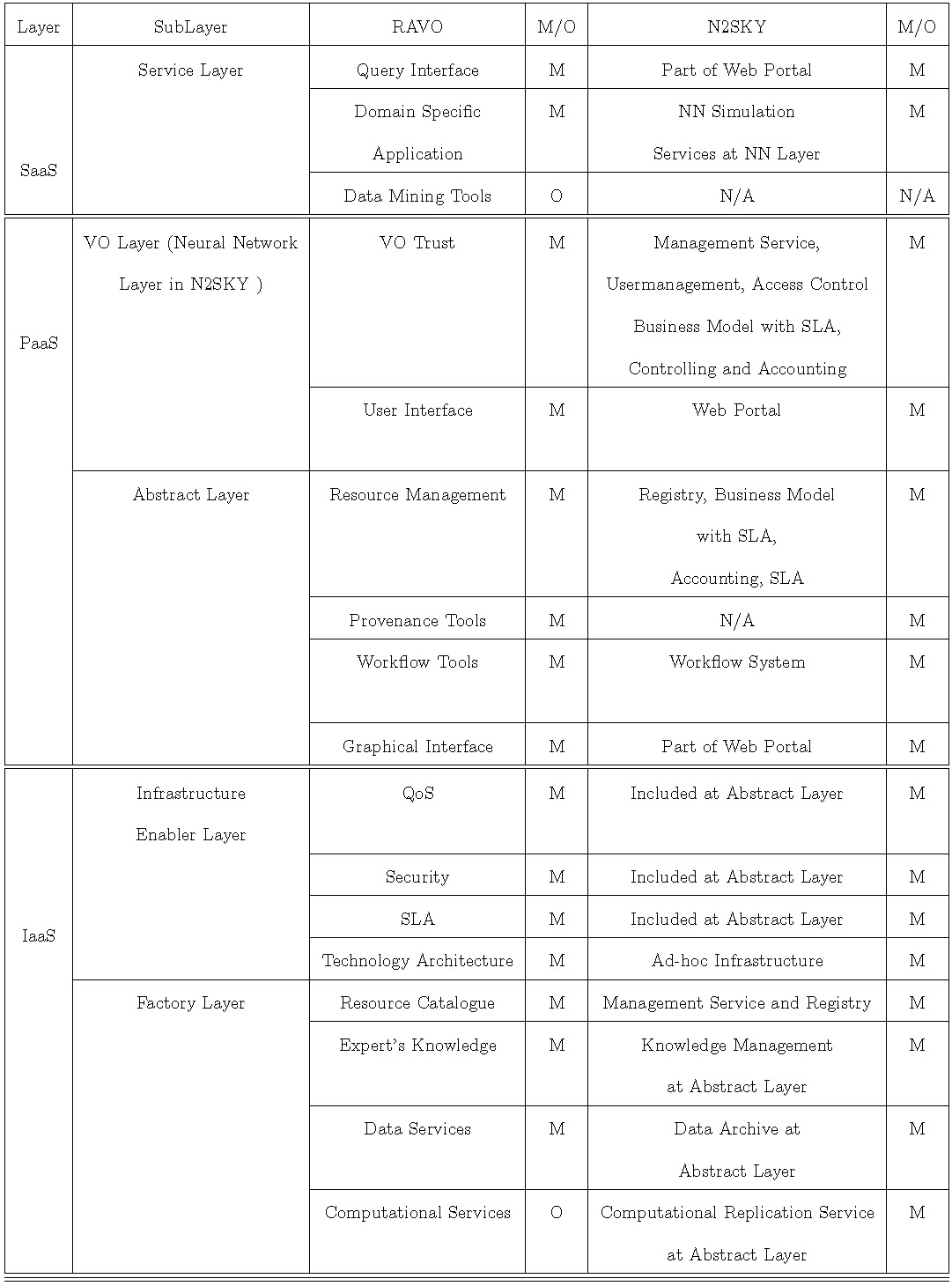}
  \caption{Comparison: RAVO vs N2Sky}
  \label{fig:comp}
  \end{center}
   \end{figure}

N2Sky is a layered architecture instantiated from RAVO. The N2Sky is shown in Figure \ref{Fig:N2S}. N2Sky is also presented as an XaaS, based on Cloud SPI model. It consists of 3 layers, namely SaaS, PaaS and IaaS. These layers have sublayers similar to RAVO. Each layer has some components which are either mandatory or optional depending upon their participation in VO. Figure \ref{Fig:CSVO} shows RAVO framework. A detailed, tabular comparison of RAVO and N2Sky components is given in Figure \ref{fig:comp}.

\subsection{Interface Specification in terms of RAVO}

 Section \ref{sec:ravo} presented interface specification for components. Here, we analyze how these interface specifications were used in N2Sky. We compare the underlying framework RAVO with its instantiation as N2Sky, in a top-down fashion. We start with SaaS layer.

\subsubsection{SaaS Layer Comparison}
SaaS layer of RAVO consists of optional and mandatory components. Choice of components and decision on their status (mandatory and optional) is open for the developers. The inclusion of components is dependent on the requirement definition by the stakeholders.

SaaS Layer has one layer, named Service layer. Here tables are included for the sake of comparison.
\begin{itemize}
\item Query Interface: RAVO proposes Query Interface as a mandatory component at Service Layer. In N2Sky, Query Interface is also included as a mandatory component.
\item Domain Specific Application (DSA): DSA is a mandatory component. N2Sky has a simulation service but at Neural Network layer (sub layer of PaaS). N2Sky includes DSA as NN specific applications. N2Sky is planned to include NN specific applications. The Simulation Service provides the creation, training and simulation of neural objects which in turn are instances of NN paradigms. Currently, Simulation Services are provided at NN Layer of N2Sky.
\item Data Mining Tools: Data mining tools are an optional component of RAVO. N2Sky has not included this option.
    \end{itemize}

N2Sky also has one layer, named Service Layer (similar to RAVO). Extended components included at Service Layer in N2Sky are:
\begin{itemize}
\item Web Portal: N2Sky Web Portal is a mandatory component.
\item Smaprtphone APP.
\item Hosted UI.
\end{itemize}

\subsubsection{PaaS Layer Comparison}
PaaS layer is composed of two layers, namely VO Layer and 2-Abstract Layer. Component Specification is detailed below.
In N2Sky PaaS consists of 3-Neural Network Layer and 2-Abstract Layer.
\paragraph{VO Layer comparison with 3-Neural Network Layer.}
In RAVO VO layer has the following components:
\begin{itemize}
\item VO Trust: Mandatory component of VO, which is responsible for enabling resources, defining policies to achieve a goal. It has several components and is extendable according to the need and requirement of stakeholders. N2Sky has distributed Trust component in to different modules. In N2Sky, Neural Network Layer has a  Management Service component to serve the purpose. Other components are available at Abstract layer namely, Business Model with SLAs and Accounting.
\item User Interface: User Interface is a mandatory component for solving problem utilizing VO PaaS utility. It provides an interface to interact with the VO. N2Sky also realizes this component as a part of Web Portal.

\end{itemize}
Extended Component of N2Sky:
\begin{itemize}
\item Hosted Component: Provides and interface for components hosting platform.
\item Simulation Service: Already described in Service Layer comparison. It is a mandatory component that is part of Neural Network Layer of N2Sky.
\end{itemize}

 \paragraph{Abstract Layer Comparison.}\label{sub:ALCompared}

 RAVO and N2Sky both have this sub layer named 2-Abstract Layer. Components of these layer in RAVO and N2Sky are compared.
\begin{itemize}
\item Resource Management: Resource Management is a mandatory component of Abstract Layer. It provides a mechanism to select and aggregate resources for a problem solving activity. Depending upon the underlying technology, VO developers can deploy different resource management tools. In N2Sky resource management is achieved via mandatory Registry component.

\item Workflow Tools: N2Sky also has a Workflow System under development.

\item Provenance Tools: Provenance Tools are proposed in RAVO but they are not included in N2Sky.
\item Graphical Interface: A mandatory components which facilitates interaction with VO easier and helps user to get results in an understandable format. It also assists user in formulating queries and browsing in VO environment. In N2Sky Graphical Interface is implemented as a Web portal described earlier.
\end{itemize}

  Extended Components supporting VO Trust (as proposed in RAVO) Functionality:
  \begin{itemize}
  \item Controlling and Accounting: This component along with SLA component serves as a Business Model. In RAVO Business Model is optional.
  \item Usermanagement.
  \item Access Control.
  \item SLA.
  \item Annotation Service.
  \item Knowledge Management: It refers to Expert's Knowledge of RAVO defined at Factory level.
  \item Component Hosting Platform.
  \end{itemize}

\subsubsection{IaaS Layer Comparison}
IaaS layer is composed of 1-Infrastructure Enabler Layer and 0-Factory Layer. This layer froms the fabric of RAVO. All the resources are available in Factory Layer and are exploited through Infrastructure Enabler Layer.

Infrastructure Enabler Layer in RAVO brings an open choice for the developers for underlying technology. QoS, Service Level Agreement (SLA), Security, Fault tolerance and Disaster management are  aspects to be considered in particular. Further extension can be done by developers.

N2Sky also have an Infrastructure Enabler Layer. It contains following components.
\begin{itemize}
\item Data Archive: Implemented as a mandatory component of N2Sky.
\item Component Replication Service.
\end{itemize}

Factory Level of RAVO is also instantiated in N2Sky. It has following components in RAVO
\begin{itemize}
\item Resource Catalogue: Resource Catalogue module is an extension of Resource Management Component. It is a mandatory components. It keeps information about resources which is of interest to VO. In N2Sky this task is achieved by Registry component.
\item Computational Services: RAVO offers Computational Services as a mandatory component. In N2Sky this component is realized by Component Replication Service. It is a mandatory component which act as N2Sky Paradigm Archive Service.
\item Data Services: This component of RAVO is realized by N2Sky as a part of Infrastructure Enabler Layer.
\item Expert's Knowledge: N2Sky implements this component of RAVO as Knowledge Management as a subcomponent of Abstract Layer.
\end{itemize}

\subsubsection{Evaluation}\label{sec:evaluation}

Based on the experiences drawn from development of N2Sky we could derive the following list of findings:

\begin{itemize}
\item RAVO provides strong theoretical grounds to clear the vision of VO developers and participants before they start building a community.
\item Requirement Analysis and Component Identification phases enable developers to list mandatory and optional components. The purpose is twofold. First, must parts of the VO are confirmed. Second, optional parts leave room for future requirements and upgrades.
\item RAVO framework is flexible and generic. Components at different layers are moved or integrated with other parts as it eases the developing process.
\item RAVO is technology independent and it gives freedom of choosing any suitable tools and programming languages.
\item RAVO emphasis on providing graphical interface to ease the end user so that they can communicate and formulate their queries easily. The interface should not be complicated that only professionals can interact.
    \item Stakeholders and their roles are important to understand. Pattern developed for RAVO are used here extensively to design a Business Model for N2Sky.
\end{itemize}

For further justification of our approach also an informal evaluation was conducted. hereby two software developers (which we name A and B), who were involved in the final development of VOCI, were interviewed.
Interview partner A is a senior researcher at our research group. His expertise include SOAs (including work on mobile devices), with a specific focus on process aware information systems. He developed a light-weight modular process engine to fully support external monitoring and intervention. He further published in the field of RESTful service description, composition and evolution. He was interviewed regarding RAVO as a staring point for developers in NN domain.

Interview partner B was a master student at University of Vienna, chose RAVO as the template to develop a cloud-based virtual organization for NNs called N2Sky. Mr. Erwin Mann has experience in implementing service-oriented architectures (SOAs), service orchestration, to create workflows and in porting such systems to cloud-based environment.
N2Sky brings together both NN paradigm developers and users who deal with problems that are beyond conventional computing possibilities. N2Sky provides a standardized description language for describing neural objects (instances of neural paradigms) called VINNSL. Furthermore N2Sky provides a business model for researchers and students but also for any interested customer. N2Sky's core process is the Simulation Service including creation, training and evaluating of neural objects in a distributed manner in the cloud.

Both researcher gave their opinion after critical analysis of RAVO. Researcher A's abstraction of RAVO in terms of ``Q\&A'' is helpful for the developers of VO in any domain. Researcher B has applied RAVO and developing an instance in the domain of NN. We deduce the following statements from this evaluation.

\begin{itemize}
\item RAVO best fits needs of community for developing a VO from scratch.
\item RAVO supports evolution of existing systems in to a VO. N2Sky is an example of such evolution.
\item RAVO is presented in a layered fashion, with a choice of mandatory and optional components. Layered approach make it easy to distribute the components in different layers and also developers are not bound to choose the exact distribution. RAVO is a flexible and extendable framework. Developer can change the components and move them to any desired layer. For example in N2Sky, components have been moved to different layer as compared to RAVO.
\item RAVO is not technology dependent. Both the researchers described their alternative choices which establishes the technological independence of RAVO.
\item Categorization of resources into logical and physical is a new dimension for VO developers. Inclusion of human expertise as a resource supports the demanding nature of problem solving ability, thereby increasing the level of trust in users.
\item RAVO presented a new concept of stakeholder, \emph{Subject}. A unique idea of how a stakeholder can become a resource in a VO. Being consumer and producer at the same time is difficult to implement. RAVO make it easier by introducing the stakeholder categorization.
\item RAVO foresees a Business Model which is introduced in N2Sky as a mandatory component. Stakeholder's roles are integrated in Business Model to set the usage and cost policy.

\end{itemize}

The interviews and the derived results are detailed in \cite{wajeehaphd12}.

\section{Conclusion}

The strong response from research community, motivated us to design a Reference Architecture for Virtual Organization (RAVO). Reference architectures are system specific and provide a level of detail required to translate the required capabilities as derived from missions, operational concepts, and operational architecture views, into projects within a capable package, which will increase system quality and decrease system development costs.

We developed a service-stack pattern to start a Virtual Organization from scratch and also presented a design blueprint for collaborative environments. Emphasis is on flexible and simple interface for interaction from the user perspective. It supports addition of tools and application to the Virtual Organization environment. We give guidelines for building a domain specific Virtual Organization for Computational Intelligence community and can extend it according to our requirements.


\begin{thebibliography}{10}

\bibitem{weishaeupl_business_2005}
T.~Weish\"{a}upl, F.~Donno, E.~Schikuta, H.~Stockinger, and H.~Wanek,
  ``Business in the grid: {BIG} project,'' in {\em Grid Economics \& Business
  Models (GECON 2005) of Global Grid Forum}, vol.~13, (Seoul, Korea), {GGF},
  2005.

\bibitem{schikuta_business_2005}
E.~Schikuta, F.~Donno, H.~Stockinger, H.~Wanek, T.~Weish\"{a}upl, E.~Vinek, and
  C.~Witzany, ``Business in the grid: Project results,'' in {\em 1st Austrian
  Grid Symposium}, (Hagenberg, Austria), {OCG}, 2005.

\bibitem{weishaeupl_towards_2004}
T.~Weish\"{a}upl and E.~Schikuta, ``Towards the merger of grid and economy,''
  in {\em International Workshop on Agents and Autonomic Computing and Grid
  Enabled Virtual Organizations {(AAC-GEVO04)} at the 3rd International
  Conference on Grid and Cooperative Computing {(GCC04)}}, vol.~3252/2004 of
  {\em Lecture Notes in Computer Science}, (Wuhan, China),
  p.~563{\textendash}570, Springer Berlin / Heidelberg, 2004.

\bibitem{weishaeupl_gset:_2006}
T.~Weish\"{a}upl, C.~Witzany, and E.~Schikuta, ``{gSET:} trust management and
  secure accounting for business in the grid,'' in {\em 6th {IEEE}
  International Symposium on Cluster Computing and the Grid {(CCGrid'06)}},
  (Singapore), p.~349{\textendash}356, {IEEE} Computer Society, 2006.

\bibitem{cs3939}
W.~Mach, B.~Pittl, and E.~Schikuta, ``A business rules driven framework for
  consumer provider contracting of web services,'' in {\em 15th International
  Conference on Information Integration and Web-based Applications \& Services
  (iiWAS2013)}, December 2013.

\bibitem{cs3713}
R.~Vigne, W.~Mach, and E.~Schikuta, ``Towards a smart webservice marketplace,''
  in {\em IEEE Conference on Business Informatics}, (USA), IEEE, 2013.

\bibitem{machschikutaIIWAS12}
W.~Mach and E.~Schikuta, ``A generic negotiation and re-negotiation framework
  for consumer-provider contracting of web services,'' in {\em 14th
  International Conference on Information Integration and Web-based
  Applications \& Services (iiWAS2012)}, (Bali, Indonesia), ACM, Dec. 2012.

\bibitem{Drucker1994}
P.~F. Drucker, {\em {Post-Capitalist} Society}.
\newblock {HarperCollins}, Nov. 2009.

\bibitem{Kesselman2008}
C.~Kesselman, I.~Foster, J.~Cummings, K.~A. Lawrence, and T.~Finholt, ``Beyond
  being there: A blueprint for advancing the design, development, and
  evaluation of virtual organizations.,'' tech. rep., May 2008.

\bibitem{WAWulf1993}
W.~Wulf, ``The collaboratory opportunity,'' {\em Science}, vol.~261, no.~5123,
  pp.~854--855, 1993.

\bibitem{Hey2005}
T.~Hey and A.~E. Trefethen, ``E-science, cyberinfrastructure and grid
  middleware services,'' {\em Science}, vol.~308, pp.~817--812, 2005.

\bibitem{OLeary2007}
M.~O'Leary and J.~Cummings, ``The spatial, temporal, and configurational
  characteristics of geographic dispersion in teams,'' {\em MIS Quarterly},
  vol.~31, pp.~433--452, March 2007.

\bibitem{Katzy2005}
B.~Katzy, C.~Zhang, and H.~Löh, ``Reference models for virtual
  organisations,'' in {\em Virtual Organizations} (L.~M. Camarinha-Matos,
  H.~Afsarmanesh, and M.~Ollus, eds.), pp.~45--58, Springer US, 2005.

\bibitem{Preece2000}
P.~Jennifer, {\em Online communities: Designing usability and supporting
  sociability}.
\newblock Chichester, England: John Wiley \& Sons Ltd, 2000.

\bibitem{Plummer2008cloud}
D.~Plummer, T.~Bittman, T.~Austin, D.~Cearley, and D.~Smith, ``Cloud computing:
  Defining and describing an emerging phenomenon,'' {\em Gartner, June},
  vol.~17, 2008.

\bibitem{schikuta_vipios_2002}
E.~Schikuta and T.~F\"{u}rle, ``{ViPIOS} islands: Utilizing {I/O} resources on
  distributed clusters,'' in {\em 15th International Conference on Parallel and
  Distributed Computing Systems {(PDCS'02)}}, (Louisville, {KY,} {USA}),
  {ISCA}, 2002.

\bibitem{schikuta_vipios:_1998}
E.~Schikuta, T.~F\"{u}rle, and H.~Wanek, ``{ViPIOS:} the vienna parallel
  {Input/Output} system,'' in {\em 4th International {Euro-Par} Conference},
  vol.~1470/1998 of {\em Lecture Notes in Computer Science}, (Southampton,
  {UK}), p.~953{\textendash}958, Springer Berlin / Heidelberg, 1998.

\bibitem{brezany_software_1996}
P.~Brezany, T.~M\"{u}ck, and E.~Schikuta, ``A software architecture for
  massively parallel input-output,'' in {\em Third International Workshop
  Applied Parallel Computing Industrial Computation and Optimization
  {(PARA'96)}} (J.~Wasniewski, J.~Dongarra, K.~Madsen, and D.~Olesen, eds.),
  vol.~1184 of {\em Lecture Notes in Computer Science}, (Lyngby, Denmark),
  p.~85{\textendash}96, Springer Berlin / Heidelberg, 1996.
\newblock 10.1007/3-540-62095-8\_10.

\bibitem{birn}
I.~Foster and C.~Kesselman, {\em {The grid: blueprint for a new computing
  infrastructure}}.
\newblock Morgan Kaufmann, 2004.

\bibitem{lead}
Accessible at : http://portal.leadproject.org/gridsphere/gridsphere.

\bibitem{nanohub}
G.~Klimeck, M.~McLennan, M.~Mannino, M.~Korkusinski, C.~Heitzinger, R.~Kennell,
  and S.~Clark, ``{NEMO 3-D and nanoHUB: Bridging Research and Education},'' in
  {\em Nanotechnology, 2006. IEEE-NANO 2006. Sixth IEEE Conference on}, vol.~2,
  pp.~441--444, IEEE, 2006.

\bibitem{cabig}
A.~C. von Eschenbach and K.~Buetow, ``{Cancer Informatics Vision: caBIG},''
  {\em Cancer informatics}, vol.~2, p.~22, 2006.

\bibitem{teragrid}
C.~Catlett, W.~Allcock, P.~Andrews, R.~Aydt, R.~Bair, N.~Balac, B.~Banister,
  T.~Barker, M.~Bartelt, P.~Beckman, {\em et~al.}, ``{Teragrid: Analysis of
  organization, system architecture, and middleware enabling new types of
  applications},'' {\em HPC and Grids in Action, Amsterdam}, 2007.

\bibitem{geon}
U.~Nambiar, B.~Ludaescher, K.~Lin, and C.~Baru, ``{The GEON portal:
  accelerating knowledge discovery in the geosciences},'' in {\em Proceedings
  of the 8th annual ACM international workshop on Web information and data
  management}, pp.~83--90, ACM, 2006.

\bibitem{earthsystemgrid}
I.~Foster, ``Service-oriented science: Scaling escience impact,'' in {\em IAT
  '06: Proceedings of the IEEE/WIC/ACM international conference on Intelligent
  Agent Technology}, (Washington, DC, USA), pp.~9--10, IEEE Computer Society,
  2006.

\bibitem{Gannon2008}
D.~Gannon, ``Building virtual organizations around super computing grids and
  clouds.'' Indiana University and Tera Grid Infrastructure Group, 2008.

\bibitem{Gerrit}
G.~Muller, ``{A Reference Architecture Primer},'' {\em Eindhoven Univ. of
  Techn., Eindhoven, White paper}, 2008.
\newblock Accessible at:
  http://www.gaudisite.nl/referencearchitectureprimerslides.pdf.

\bibitem{nexof}
``Nexof ra.'' Accessible at: http://www.nexof-ra.eu/.

\bibitem{Altafwcci}
A.~A. Huqqani, P.~Beran, X.~Li, and E.~S. ., ``N2cloud: Cloud based neural
  network simulation application,'' in {\em In Proceedings of the International
  Joint Conference on Neural Networks 2010 (WCCI 2010), Barcelona, Spain},
  2010.

\bibitem{Jen2000}
L.~Jen and Y.~Lee, ``Working group. ieee recommended practice for architectural
  description of software-intensive systems,'' in {\em IEEE Architecture},
  Citeseer, 2000.

\bibitem{Rinder2008}
S.~Rinderle-Ma and M.~Reichert, ``Managing the life cycle of access rules in
  ceosis,'' in {\em Enterprise Distributed Object Computing Conference, 2008.
  EDOC'08. 12th International IEEE}, pp.~257--266, IEEE, 2008.

\bibitem{Rinder2004}
S.~Rinderle, M.~Reichert, and P.~Dadam, ``Correctness criteria for dynamic
  changes in workflow systems--a survey,'' {\em Data \& Knowledge Engineering},
  vol.~50, no.~1, pp.~9--34, 2004.

\bibitem{FProcess2008}
H.~Schonenberg, R.~Mans, N.~Russell, N.~Mulyar, and W.~Aalst, ``Process
  flexibility: A survey of contemporary approaches,'' {\em Advances in
  Enterprise Engineering I}, pp.~16--30, 2008.

\bibitem{Hasan2007}
R.~Hasan, R.~Sion, and M.~Winslett, ``Introducing secure provenance: problems
  and challenges,'' in {\em Proceedings of the 2007 ACM workshop on Storage
  security and survivability}, StorageSS '07, (New York, NY, USA), pp.~13--18,
  ACM, 2007.

\bibitem{haq_sla_2010}
I.~U. Haq, E.~Schikuta, I.~Brandic, A.~Paschke, and H.~Boley, ``{SLA}
  validation of service value chains,'' in {\em 9th International Conference on
  Grid and Cloud Computing {(GCC'10)}}, (Nanjing, Jiangsu, China),
  p.~308{\textendash}313, {IEEE} Computer Society, 2010.

\bibitem{haq_aggregation_2010}
I.~U. Haq and E.~Schikuta, ``Aggregation patterns of service level
  agreements,'' in {\em Frontiers of Information Technology {(FIT'10)}},
  (Islamabad, Pakistan), {ACM}, 2010.

\bibitem{haq_sla_2010-1}
I.~U. Haq, I.~Brandic, and E.~Schikuta, ``{SLA} validation in layered cloud
  infrastructures,'' in {\em Economics of Grids, Clouds, Systems, and Services,
  7th International Workshop, {GECON'10}}, vol.~6296 of {\em Lecture Notes in
  Computer Science}, (Ischia, Italy), p.~153{\textendash}164, Springer Berlin /
  Heidelberg, 2010.

\bibitem{haq_conceptual_2009}
I.~U. Haq, A.~A. Huqqani, and E.~Schikuta, ``A conceptual model for aggregation
  and validation of {SLAs} in business value networks,'' in {\em 3rd
  International Conference on Adaptive Business Information Systems
  {(ABIS'09)}}, (Leipzig, Germany), Mar. 2009.

\bibitem{haq_rule-based_2009}
I.~U.~U. Haq, A.~Paschke, E.~Schikuta, and H.~Boley, ``{Rule-Based} workflow
  validation of hierarchical service level agreements,'' in {\em Workshops at
  the Grid and Pervasive Computing Conference {(GPC'09)}},
  p.~96{\textendash}103, {IEEE} Computer Society, 2009.

\bibitem{haq_aggregating_2009}
I.~U. Haq, A.~Huqqani, and E.~Schikuta, ``Aggregating hierarchical service
  level agreements in business value networks,'' in {\em 7th International
  Conference on Business Process Management {(BPM'09)}}, (Ulm, Germany),
  p.~176{\textendash}192, {Springer-Verlag}, 2009.

\bibitem{mann13}
E.~Mann, ``N2sky - a cloud-based artificial neural network resource,'' master
  thesis, Faculty of Computer Science, university of Vienna, A-1090 Vienna,
  Austria, 2013.

\bibitem{wajeehaphd12}
W.~Khalil, {\em Reference Architecture for Virtual Organization}.
\newblock Ph.d. thesis, Faculty of Computer Science, university of Vienna,
  A-1090 Vienna, Austria, 2013.

\end{thebibliography}

\end{document}